\documentclass[12pt]{article}

\usepackage{amsmath}
\usepackage{graphicx}

\newcommand{\ket}[1]{| #1 \rangle}
\newcommand{\bra}[1]{\langle #1 |}
\newcommand{\hcs}[1]{#1^\dagger #1}

\begin{document}

\begin{center}
{\Large\bf Reversible quantum measurement
with arbitrary spins}
\vskip .6 cm
Hiroaki Terashima$^{1,2}$ and Masahito Ueda$^{1,2}$
\vskip .4 cm
{\it $^1$Department of Physics, Tokyo Institute of Technology,\\
Tokyo 152-8551, Japan} \\
{\it $^2$CREST, Japan Science and Technology Corporation (JST),\\
Saitama 332-0012, Japan}
\vskip .6 cm
\end{center}

\begin{abstract}
We propose a physically reversible
quantum measurement of an arbitrary spin-$s$ system
using a spin-$j$ probe via an Ising interaction.
In the case of a spin-$1/2$ system ($s=1/2$),
we explicitly construct a reversing measurement
and evaluate the degree of reversibility
in terms of fidelity.
The recovery of the measured state is pronounced
when the probe has a high spin ($j>1/2$),
because the fidelity changes drastically during
the reversible measurement and the reversing measurement.
We also show that
the reversing measurement scheme for a spin-$1/2$ system 
can serve as an experimentally feasible approximate
reversing measurement for a high-spin system ($s>1/2$).
If the interaction is sufficiently weak,
the reversing measurement can recover a cat state
almost deterministically
in spite of there being a large fidelity change.
\end{abstract}

\begin{flushleft}
{\footnotesize
}
\end{flushleft}

\section{Introduction}
Quantum measurements are widely believed
to have intrinsic irreversibility,
since they play different roles with respect to
the past and future of
the measured system~\cite{LanLif77}.
With respect to the past,
a quantum measurement verifies the predicted probabilities
for possible outcomes.
With respect to the future,
a measurement brings about a new quantum state
via nonunitary state reduction.
However, as shown in Ref.~\cite{UedKit92},
a quantum measurement is not necessarily irreversible.
A quantum measurement is said to
be logically reversible~\cite{UedKit92,UeImNa96}
if the premeasurement state can be calculated from
the postmeasurement state and the outcome of the measurement.
This means that all the information about
the premeasurement state is
preserved during the measurement.
A quantum measurement is said to
be physically reversible~\cite{UeImNa96,Ueda97}
if the premeasurement state can be
recovered from the postmeasurement state
by means of a second measurement,
referred to as a reversing measurement,
with a nonzero probability.
In this case, not only is
the information about the system preserved
during the measurement process,
but the original state can be restored
by means of a physical process.

Some measurements are known to be
logically reversible~\cite{UedKit92,Imamog93,Royer94}.
Royer~\cite{Royer94} proposed
a physically reversible quantum measurement of
a spin-$1/2$ system using a spin-$1/2$ probe
in an attempt to completely determine the unknown
quantum state of a single system
(see, however, Erratum of Ref.~\cite{Royer94}).
In the context of quantum computation~\cite{NieChu00},
the reversing measurement has been discussed
for reducing the qubit overhead
in quantum error correction~\cite{KoaUed99} and
for improving the probability of
successful nonunitary gate operation
in a nonunitary quantum circuit~\cite{TerUed03}.
As an important step toward the experimental realization
of a reversible measurement,
a photodetection scheme that satisfies a necessary condition
for logical reversibility (``sensitivity to vacuum fluctuations'')
has recently been demonstrated~\cite{UNSTN04}
using a stimulated parametric down-conversion process.

In this paper, we propose
a scheme for making a physically reversible quantum measurement
that is experimentally feasible
in view of recent advances in experimental
techniques~\cite{KMJYEB99,GeStMa04}.
Our model consists of two arbitrary spin systems
(a measured system and a probe system)
interacting via an Ising Hamiltonian.
Since spin can describe diverse physical systems
(e.g., the real spin of particles, collective two-level systems,
Cooper pairs, interferometers, and Josephson junctions),
our model can be used to implement both physically
reversible measurements and reversing measurements
in such diverse systems.
We explicitly construct
a reversing measurement for our model,
in which quantitative analysis is performed
in terms of fidelity~\cite{NieChu00}.
When the probe system has a high spin,
the fidelity changes drastically
in both the reversible measurement and the reversing measurement.
The high-spin probe thus makes the recovery of
the measured state more pronounced than for the spin-$1/2$ model,
though at the cost of
decreasing the probability of successful recovery.

To clarify what kind of irreversibility is at issue,
we here review a projective measurement~\cite{vonNeu96},
which is often used to describe measurement processes
in quantum theory.
Let $\hat{O}$ be a measured observable,
whose eigenstate with eigenvalue $m$ is denoted by $\ket{m}$.
The observable $\hat{O}$ can then be decomposed
as $\sum_m m\hat{P}_m$,
where $\hat{P}_m$ is the projector $\ket{m}\bra{m}$.
From the completeness condition,
the projectors $\{\hat{P}_m\}$ satisfy
\begin{equation}
\sum_m\hat{P}_m=\hat{I},
\end{equation}
with $\hat{I}$ being the identity operator.
Suppose that the measured system is initially in a state $\ket{\psi}$.
The projective measurement with respect to $\{\hat{P}_m\}$
yields an outcome $m$ with probability
\begin{equation}
 p_m=\bra{\psi}\hat{P}_m\ket{\psi},
\end{equation}
and then causes a state reduction of the measured system to
\begin{equation}
\ket{\psi_m}=\frac{1}{\sqrt{p_m}}\hat{P}_m\ket{\psi}.
\end{equation}
Clearly, the projective measurement is irreversible
in the sense that we cannot recover
the premeasurement state $\ket{\psi}$
from the postmeasurement state $\ket{\psi_m}$,
unless we \textit{a priori} know the former state.
This is because the information about the states
orthogonal to $\hat{P}_m$ is completely lost
during the measurement.
One might think that
any quantum measurement has this type of irreversibility,
since quantum measurement entails
a nonunitary state reduction
associated with information readout.
However, there exist quantum measurements
that are logically reversible in spite of
nonunitary state reduction~\cite{UedKit92,Imamog93,Royer94}.

To formulate the conditions for logical reversibility,
we adopt a general formulation of
quantum measurement~\cite{DavLew70,NieChu00}, in which
a quantum measurement is described by a set of
measurement operators $\{\hat{M}_m\}$
that satisfies
\begin{equation}
\sum_m\hcs{\hat{M}_m}=\hat{I}.
\end{equation}
If the measured system is in a state $\ket{\psi}$,
the general measurement with respect to $\{\hat{M}_m\}$
yields an outcome $m$ with probability
\begin{equation}
 p_m=\bra{\psi}\hcs{\hat{M}_m}\ket{\psi},
\end{equation}
and then causes a state reduction of the measured system to
\begin{equation}
\ket{\psi_m}=\frac{1}{\sqrt{p_m}}\hat{M}_m\ket{\psi}.
\end{equation}
Note that this state change depends on the outcome $m$.
The general measurement can be simulated by
a projective measurement with the help of a measurement probe,
even though the projective measurement
is a special case of
the general measurement ($\hat{M}_m=\hat{P}_m$).
The necessary and sufficient condition
for logical reversibility is
$\hat{M}_m\ket{\psi}\neq 0$ for any $\ket{\psi}$
in the Hilbert space~\cite{UeImNa96}.
In other words, the measurement must
respond to any input state so that
no possibility of the premeasurement state
is excluded by any outcome of the measurement.
For example, usual photon counting~\cite{UeImOg90} is
logically irreversible because the detection of a photon
excludes the possibility that the premeasurement state
is the vacuum state.
On the other hand,
the necessary and sufficient condition
for physical reversibility is that $\hat{M}_m$ has
a bounded left inverse~\cite{UeImNa96,Ueda97}.
Thus physical reversibility implies
logical reversibility, but not vice versa.
An important special case is that of
a finite-dimensional Hilbert space, where
physical reversibility
is equivalent to logical reversibility.
However, in an infinite-dimensional Hilbert space,
there exist logically reversible
yet physically irreversible measurements~\cite{UeImNa96}
such as quantum counting~\cite{UedKit92}.

A different type of reversibility
is discussed in Refs.~\cite{MabZol96,NieCav97}.
A quantum measurement is said to
be unitarily reversible if the premeasurement state
can be recovered by a reversing unitary operation
on the postmeasurement state.
In this case,
although successful reversal occurs with unit probability
owing to the unitarity,
it is essential that the premeasurement state lie within
a certain subspace of the entire Hilbert space.
Since the subspace is chosen so that
the probability of each measurement outcome is the same
for all states in the subspace,
no information about the premeasurement state can be obtained
from the unitarily reversible quantum measurement~\cite{NieCav97}.

This paper is organized as follows.
Section~\ref{sec:measure} formulates
a physically reversible quantum measurement of
a spin-$s$ system using a spin-$j$ probe.
Section~\ref{sec:rev-mea1} explicitly constructs
the reversing measurement
for the case of a measured system with $s=1/2$,
focusing on the effect of a high-spin probe ($j>1/2$).
Section~\ref{sec:rev-mea2} describes
two approximate schemes of the reversing measurement
for the case of measured systems with $s>1/2$:
one in which the measured system is initially
in a two-dimensional subspace
and the other in which
the interaction is sufficiently weak.
Section~\ref{sec:exp} discusses a possible experimental
situation using an ensemble of atoms as a measured system
and two-mode photons as a probe system.
Section~\ref{sec:conclude} summarizes our results.
Throughout this paper, we refer to the measured system
and the probe system simply as
\emph{system} and \emph{probe}, respectively.

\section{\label{sec:measure}Reversible Spin Measurement}
First, we formulate a quantum measurement of a spin-$s$ system
described by spin operators
$\{\hat{S}_{x},\hat{S}_{y},\hat{S}_{z}\}$.
These operators obey the commutation relations
\begin{equation}
[ \hat{S}_{i},\hat{S}_{j} ] = i \epsilon_{ijk}\hbar \hat{S}_{k},
\end{equation}
where the indices $i,j,k$ denote $x,y,z$ and
$\epsilon_{ijk}$ is the Levi-Civita symbol.
The Hilbert space of this system is spanned by
the eigenstates of $\hat{S}_{z}$,
\begin{equation}
\hat{S}_{z}\ket{\sigma}_q= \sigma\hbar\,\ket{\sigma}_q,
\end{equation}
where $\sigma=s,s-1,\ldots,-s+1,-s$.
Using these states, the state to be measured is written as
\begin{equation}
\ket{\psi}_q=\sum_{\sigma} c_\sigma \ket{\sigma}_q
\label{eq:psi}
\end{equation}
with the normalization condition
\begin{equation}
\sum_{\sigma} |c_\sigma|^2=1.
\label{eq:cnorm}
\end{equation}
It should be emphasized that
the coefficients $\{c_\sigma\}$ are unknown,
since it is assumed that we have no \textit{a priori}
information about the measured state $\ket{\psi}_q$.
The measured system is assumed to be in a pure state
as in Eq.~(\ref{eq:psi});
a mixed initial state of the system makes no difference
in constructing a reversing measurement.

To measure the spin state of the system,
we introduce a probe with spin $j$.
The probe is described by spin operators
$\{\hat{J}_{x},\hat{J}_{y},\hat{J}_{z}\}$
satisfying the commutation relations
\begin{equation}
[ \hat{J}_{i},\hat{J}_{j} ] = i\epsilon_{ijk}\hbar \hat{J}_{k}.
\label{eq:probeso}
\end{equation}
The Hilbert space of this system is also spanned by
the eigenstates of $\hat{J}_{z}$,
\begin{equation}
\hat{J}_{z}\ket{m}_p= m\hbar\,\ket{m}_p,
\end{equation}
where $m=j,j-1,\ldots,-j+1,-j$.

We prepare the probe in a state
\begin{align}
\ket{\theta,\phi}_p &=
\exp\left(-\frac{i}{\hbar}\hat{J}_{z}\phi\right)
\exp\left(-\frac{i}{\hbar}\hat{J}_{y}\theta\right)\ket{j}_p \notag \\
 &= \sum_m e^{-im\phi} d_{mj}^{(j)}(\theta)\, \ket{m}_p,
\label{eq:probinit}
\end{align}
where $d_{m'm}^{(j)}(\theta)$ is defined by~\cite{Sakura94}
\begin{align}
 d_{m'm}^{(j)}(\theta)&\equiv {}_p\bra{m'}
  \exp\left(-\frac{i}{\hbar}\hat{J}_{y}\theta\right)\ket{m}_p \notag \\
 &= \sum_{\substack{0\le k \le j+m \\ m-m'\le k \le j-m' }}
\frac{\sqrt{(j+m)!(j-m)!(j+m')!(j-m')!}}%
{(j+m-k)!k!(j-k-m')!(k-m+m')!} \notag \\
 &\qquad {}\times  (-1)^{k-m+m'}
     \left(\cos\frac{\theta}{2}\right)^{2j-2k+m-m'}
     \left(\sin\frac{\theta}{2}\right)^{2k-m+m'}.
\end{align}
We assume that the interaction between the system and the probe
is of an Ising type,
\begin{equation}
   H=\alpha \hat{J}_{z}\hat{S}_{z},
\end{equation}
where $\alpha$ is a real constant.
This $\hat{J}_{z}\hat{S}_{z}$-type
interaction has direct relevance to experimental situations
in Refs.~\cite{HapMat67,KuBiMa98,THTTIY99,TITKYT05,KMJYEB99}.
The interaction between the system and
the probe gives rise to a unitary transformation,
\begin{equation}
 \hat{U}_i=\exp\left(-\frac{2ig}{\hbar^2}\hat{J}_{z}
             \hat{S}_{z}\right),
\label{eq:interaction}
\end{equation}
on the combined system,
where $g\equiv\alpha t\hbar/2$ is the effective strength of
the interaction.

After the interaction, the unitary operator
\begin{equation}
 \hat{U}_p=\exp\left(-\frac{i}{\hbar}\hat{J}_{y}\frac{\pi}{2}\right)
\label{eq:pulse}
\end{equation}
is applied to the probe.
The state of the whole system then becomes
\begin{equation}
 \hat{U}_p\hat{U}_i\ket{\theta,\phi}_p\ket{\psi}_q
 = \sum_{m',\sigma} a_{m'\sigma}^{(j)}(\theta,\phi)\,
  c_\sigma\ket{m'}_p\ket{\sigma}_q,
\end{equation}
where $a_{m'\sigma}^{(j)}(\theta,\phi)$ is given by
\begin{align}
 a_{m'\sigma}^{(j)}(\theta,\phi) &\equiv
  \sum_{m} e^{-im(2g\sigma+\phi)} d_{mj}^{(j)}(\theta)\,
     d_{m'm}^{(j)}\left(\frac{\pi}{2}\right) \notag \\
  &= \frac{1}{2^j}
      \sqrt{\frac{(2j)!}{(j+m')!(j-m')!}}  \notag \\
  &  \qquad{}\times
     \left( e^{-\frac{i}{2}(2g\sigma+\phi)}\cos\frac{\theta}{2}
     +e^{\frac{i}{2}(2g\sigma+\phi)}
     \sin\frac{\theta}{2} \right)^{j-m'} \notag \\
  &  \qquad{}\times
     \left( e^{-\frac{i}{2}(2g\sigma+\phi)}\cos\frac{\theta}{2}
     -e^{\frac{i}{2}(2g\sigma+\phi)}
     \sin\frac{\theta}{2} \right)^{j+m'}.
\end{align}
Note that $|a_{m'\sigma}^{(j)}(\theta,\phi)|^2$ is
a binomial distribution as a function of $m'$
\begin{figure}
\begin{center}
\includegraphics[scale=0.75]{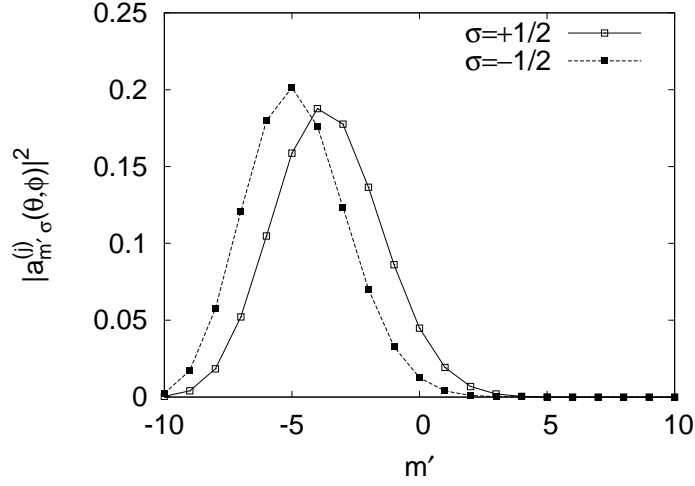}
\end{center}
\caption{\label{fig1}$|a_{m'\sigma}^{(j)}(\theta,\phi)|^2$
as a function of $m'$
($\sigma=\pm1/2$, $j=10$, $g=0.25$,
$\theta=\pi/6$, $\phi=\pi/6$).}
\end{figure}
(see Fig.~\ref{fig1}):
\begin{align}
 |a_{m'\sigma}^{(j)}(\theta,\phi)|^2  &=
   \frac{(2j)!}{(j+m')!(j-m')!}  \notag \\
  &  \qquad{}\times
   \left[\frac{1+\chi_\sigma(\theta,\phi)}{2}\right]^{j-m'}
   \left[\frac{1-\chi_\sigma(\theta,\phi)}{2}\right]^{j+m'},
\label{eq:asqr}
\end{align}
where
\begin{equation}
 \chi_\sigma(\theta,\phi)
\equiv \sin\theta\cos(2g\sigma+\phi).
\end{equation}
We thus obtain the normalization condition
\begin{equation}
\sum_{m'} |a_{m'\sigma}^{(j)}(\theta,\phi)|^2=1.
\label{eq:anorm}
\end{equation}
The mean and variance
of this distribution are given by
\begin{align}
\mu_\sigma(\theta,\phi) &\equiv
 \sum_{m'} m'|a_{m'\sigma}^{(j)}(\theta,\phi)|^2 \notag \\
 &= -j\chi_\sigma(\theta,\phi) \label{eq:bimean}
\end{align}
and
\begin{align}
 \nu_\sigma(\theta,\phi)  &\equiv \sum_{m'}
\left(m'-\mu_\sigma(\theta,\phi)\right)^2
 |a_{m'\sigma}^{(j)}(\theta,\phi)|^2 \notag \\
&= j \left[
\frac{1-\chi_\sigma(\theta,\phi)^2}{2}\right],
\end{align}
respectively.
The central limit theorem states that as $j$ increases,
the binomial distribution becomes close to
a normal distribution with the mean and variance unaltered.
Thus, for large $j$, we can approximate
the distribution as
\begin{equation}
|a_{m'\sigma}^{(j)}(\theta,\phi)|^2 \sim
\frac{1}{\sqrt{2\pi\nu_\sigma(\theta,\phi)}}
\exp\left[-\frac{(m'-\mu_\sigma(\theta,\phi))^2}%
{2\nu_\sigma(\theta,\phi)}\right].
\label{eq:clt}
\end{equation}

We finally perform a projective measurement on
the probe variable $\hat{J}_{z}/\hbar$
and obtain the measurement outcome $m$ ($=j,j-1,\ldots,-j+1,-j$).
Alternatively,
we can perform the projective measurement
of $-\hat{J}_{x}/\hbar$ without the unitary operator $\hat{U}_{p}$
in Eq.~(\ref{eq:pulse}).
Since the probability for outcome $m$ is
\begin{equation}
p_m=\sum_{\sigma}
|a_{m\sigma}^{(j)}(\theta,\phi)|^2 |c_\sigma|^2,
\label{eq:pm}
\end{equation}
we can obtain information
about the initial state (\ref{eq:psi}) of the system
from this measurement
through the dependence of $p_m$ on $c_\sigma$.
However, if $|a_{m\sigma}^{(j)}(\theta,\phi)|^2$
does not depend on $\sigma$,
the probability
$p_m$ does not depend on $c_\sigma$
because of the normalization condition (\ref{eq:cnorm}).
Therefore, to obtain information about the measured state,
the initial probe state $\ket{\theta,\phi}_p$
and the strength of the interaction $g$ must satisfy
\begin{align}
 & \sin\theta \neq 0, \notag \\
 & \sin g \neq 0,    \label{eq:meacond}\\
 & \sin\left[\left(2s-1\right)g+\phi\right]\neq 0, \notag
\end{align}
according to Eq.~(\ref{eq:asqr}),
where the last condition is required if $s=1/2$,
or if $s>1/2$ and $\cos g=0$.
From Eqs.~(\ref{eq:cnorm}) and (\ref{eq:anorm}),
it is easy to see that
the total probability is
\begin{equation}
 \sum_m p_m=1.
\end{equation}
Using Eq.~(\ref{eq:bimean}),
the expected value of $m$ is given by
\begin{equation}
\overline{m}\equiv \sum_m m p_m
=-j\sum_{\sigma}\chi_\sigma(\theta,\phi)|c_\sigma|^2.
\end{equation}
The measurement process causes a nonunitary
state reduction of the measured system.
Corresponding to the outcome $m$,
the state of the system becomes
\begin{equation}
\ket{\psi_m}_q=\frac{1}{\sqrt{p_m}}
\sum_{\sigma} a_{m\sigma}^{(j)}(\theta,\phi)\,
c_\sigma \ket{\sigma}_q
\label{eq:psim}
\end{equation}
and its fidelity with the premeasurement state
decreases to
\begin{equation}
F_m =\bigl|{}_q\langle\psi |\psi_m \rangle_q\bigr|
=\frac{1}{\sqrt{p_m}} \left|\sum_{\sigma}
a_{m\sigma}^{(j)}(\theta,\phi)\,|c_\sigma|^2\right|.
\label{eq:fm}
\end{equation}

We can describe this measurement process
by a set of measurement operators,
as in the general quantum
measurement~\cite{DavLew70,NieChu00}.
Let $\hat{T}_m(\theta,\phi)$ be
the measurement operator for outcome $m$.
Since the probability (\ref{eq:pm}) and
postmeasurement state (\ref{eq:psim})
are expressed as
\begin{align}
  p_m &= {}_q\bra{\psi}\,
\hat{T}_m^\dagger(\theta,\phi)\,
\hat{T}_m(\theta,\phi)\,\ket{\psi}_q,  \\
  \ket{\psi_m}_q &= \frac{1}{\sqrt{p_m}}\,
\hat{T}_m(\theta,\phi)\,\ket{\psi}_q,
\end{align}
the explicit form of $\hat{T}_m(\theta,\phi)$
is given by
\begin{equation}
\hat{T}_m(\theta,\phi)=\sum_\sigma
  a_{m\sigma}^{(j)}(\theta,\phi)\,
\ket{\sigma}_q{}_q\bra{\sigma}.
\label{eq:meaop}
\end{equation}
From Eq.~(\ref{eq:anorm}), we can confirm that
\begin{equation}
\sum_m \hat{T}_m^\dagger(\theta,\phi)\,
\hat{T}_m(\theta,\phi)=\hat{I}.
\end{equation}
This measurement does not disturb
the eigenstates of $\hat{S}_{z}$
owing to the commutation relation
\begin{equation}
[ \hat{S}_{z}, \hat{T}_m(\theta,\phi) ]=0.
\label{eq:qnd}
\end{equation}

The measurement $\{\hat{T}_m(\theta,\phi)\}$ is
logically reversible~\cite{UedKit92,UeImNa96}
if $\hat{T}_m(\theta,\phi)\ket{\psi}_q\neq 0$
for any $\ket{\psi}_q$,
or equivalently if
$a_{m\sigma}^{(j)}(\theta,\phi)\neq0$ for any $\sigma$.
This condition requires
the initial probe state $\ket{\theta,\phi}_p$
and the strength of the interaction $g$ to satisfy
\begin{equation}
\sin\theta\neq \pm1 \quad\text{or}\quad
\cos(2g\sigma+\phi)\neq \pm 1
\label{eq:revcond}
\end{equation}
for $\sigma=s,s-1,\ldots,-s+1,-s$.
When these conditions are satisfied,
the measurement $\{\hat{T}_m(\theta,\phi)\}$
is physically reversible as well,
since $\hat{T}_m(\theta,\phi)$ has a bounded left inverse.
This implies that there exists another measurement that can
recover the \emph{unknown} premeasurement state (\ref{eq:psi})
from the postmeasurement state (\ref{eq:psim})
with a nonzero probability.
We explicitly construct such reversing measurements
in the following sections.
Note, however, that the measurement $\{\hat{T}_m(\theta,\phi)\}$
is not unitarily reversible~\cite{MabZol96,NieCav97}
if condition (\ref{eq:meacond}) is satisfied.
This is because we have obtained some information about
the measured state from the measurement outcome
via the probability that depends on
the measured state~\cite{NieCav97}.
Therefore, there is no unitary operation that can
recover the premeasurement state from the postmeasurement state.

\section{\label{sec:rev-mea1}%
Reversing Measurement on a Spin-$1/2$ \\
System ($s=1/2$)}

\subsection{Scheme}
We consider a reversing measurement of
a physically reversible measurement
$\{\hat{T}_m(\theta,\phi)\}$ for the case of
a measured system with $s=1/2$, where
the measurement operator $\hat{T}_m(\theta,\phi)$
is in the basis $\{\,\ket{1/2}_q,\ket{-1/2}_q\}$
represented by a diagonal $2\times2$ matrix as
\begin{equation}
{}_q\bra{\sigma'}\, \hat{T}_m(\theta,\phi)\,\ket{\sigma}_q
  =\left(\begin{array}{cc}
    a_{m,\frac{1}{2}}^{(j)}(\theta,\phi) & 0 \\
    0 & a_{m,-\frac{1}{2}}^{(j)}(\theta,\phi)
    \end{array}\right).
\label{eq:halfspinop}
\end{equation}

\begin{figure}
\begin{center}
\includegraphics[scale=0.57]{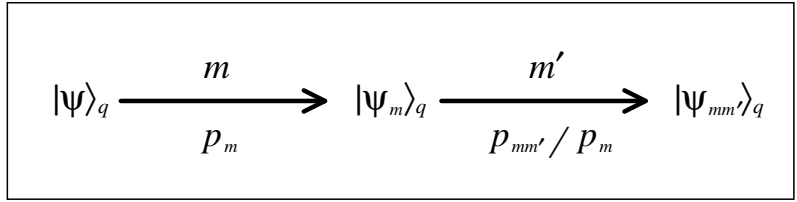}
\end{center}
\caption{\label{fig2}Transitions of the measured state
by successive measurements $\{\hat{T}_m(\theta,\phi)\}$
and $\{\hat{T}_m(\pi-\theta,\pi-\phi)\}$.
The first measurement on the state $\ket{\psi}_q$
yields an outcome $m$ ($=j,j-1,\ldots,-j+1,-j$) with
probability $p_m$, causing a state reduction to $\ket{\psi_m}_q$.
The second measurement on $\ket{\psi_m}_q$ then
yields an outcome $m'$ with conditional probability $p_{mm'}/p_m$,
causing a state reduction to $\ket{\psi_{mm'}}_q$.
}
\end{figure}

Suppose that a second measurement
$\{\hat{T}_m(\pi-\theta,\pi-\phi)\}$ is performed
on the postmeasurement state (\ref{eq:psim})
and that an outcome $m'$ ($=j,j-1,\ldots,-j+1,-j$)
is obtained, as illustrated in Fig.~\ref{fig2}.
Using
\begin{equation}
 a_{m'\sigma}^{(j)}(\pi-\theta,\pi-\phi)
  =e^{-im'\pi}a_{-m',-\sigma}^{(j)}(\theta,\phi),
\label{eq:asymm}
\end{equation}
the measurement operator
$\hat{T}_{m'}(\pi-\theta,\pi-\phi)$ is represented by
\begin{equation}
{}_q\bra{\sigma'}\, \hat{T}_{m'}(\pi-\theta,\pi-\phi)\,\ket{\sigma}_q
= e^{-im'\pi} \left(\begin{array}{cc}
    a_{-m',-\frac{1}{2}}^{(j)}(\theta,\phi) & 0 \\
    0 & a_{-m',\frac{1}{2}}^{(j)}(\theta,\phi)
    \end{array}\right).
\label{eq:halfspinop2}
\end{equation}
The state of the system thus becomes
\begin{equation}
\ket{\psi_{mm'}}_q=\frac{e^{-im'\pi}}{\sqrt{p_{mm'}}}
\sum_{\sigma=\pm\frac{1}{2}} a_{-m',-\sigma}^{(j)}(\theta,\phi)\,
 a_{m\sigma}^{(j)}(\theta,\phi)\,c_\sigma \ket{\sigma}_q,
\label{eq:psimm}
\end{equation}
where
\begin{equation}
p_{mm'}=\sum_{\sigma=\pm\frac{1}{2}}
|a_{-m',-\sigma}^{(j)}(\theta,\phi)\,
 a_{m\sigma}^{(j)}(\theta,\phi)|^2 |c_\sigma|^2
\label{eq:pmm}
\end{equation}
is the joint probability of obtaining the outcomes
$m$ for the first measurement and $m'$ for the second measurement.
The expected values of $m$ and $m'$ are given by
\begin{align}
\overline{m} &= \sum_{m,m'} m \,p_{mm'} \notag \\
 &= -j\left(\chi_{\frac{1}{2}}(\theta,\phi)\,|c_{\frac{1}{2}}|^2
+\chi_{-\frac{1}{2}}(\theta,\phi)\,|c_{-\frac{1}{2}}|^2\right), \\
\overline{m'} &= \sum_{m,m'} m'\, p_{mm'} \notag \\
 &= +j\left(\chi_{-\frac{1}{2}}(\theta,\phi)\,|c_{\frac{1}{2}}|^2
+\chi_{\frac{1}{2}}(\theta,\phi)\,|c_{-\frac{1}{2}}|^2\right),
\end{align}
respectively.
Therefore, as a function of $m$ and $m'$,
the joint probability $p_{mm'}$ has two peaks at
\begin{equation}
(m,m')=\left(-j\chi_{\pm\frac{1}{2}}(\theta,\phi),
+j\chi_{\mp\frac{1}{2}}(\theta,\phi)\right),
\label{eq:twopeak}
\end{equation}
where the heights of the peaks depend on
the values of $|c_{1/2}|^2$ and $|c_{-1/2}|^2$.

An interesting case of recovery of the measured state
occurs when the outcome of the second measurement is
the negative of the first one (i.e., $m'=-m$).
Since $a_{m,-\sigma}^{(j)}(\theta,\phi)
 a_{m\sigma}^{(j)}(\theta,\phi)$ does not depend on
$\sigma$ ($=\pm1/2$),
the final state (\ref{eq:psimm}) with $m'=-m$
is identical to the original state (\ref{eq:psi})
except for an overall phase factor,
\begin{equation}
\ket{\psi_{m,-m}}_q =e^{i\alpha}
\sum_{\sigma=\pm\frac{1}{2}}c_\sigma \ket{\sigma}_q,
\label{eq:revpsi}
\end{equation}
where
\begin{equation}
e^{i\alpha}\equiv e^{im\pi}
\frac{a_{m,-\frac{1}{2}}^{(j)}(\theta,\phi)\,%
a_{m,\frac{1}{2}}^{(j)}(\theta,\phi)}%
{|a_{m,-\frac{1}{2}}^{(j)}(\theta,\phi)\,%
a_{m,\frac{1}{2}}^{(j)}(\theta,\phi)|}.
\end{equation}
Therefore, the second measurement $\{\hat{T}_m(\pi-\theta,\pi-\phi)\}$
is a reversing measurement of
the first measurement $\{\hat{T}_m(\theta,\phi)\}$.
Here, the state recovery results from the identity
\begin{equation}
\hat{T}_{-m}(\pi-\theta,\pi-\phi)\,\hat{T}_m(\theta,\phi)
=\left[e^{im\pi}a_{m,-\frac{1}{2}}^{(j)}(\theta,\phi)\,
 a_{m,\frac{1}{2}}^{(j)}(\theta,\phi)\right]\hat{I},
\label{eq:proto}
\end{equation}
which implies that $\hat{T}_{-m}(\pi-\theta,\pi-\phi)$
is proportional to the inverse of $\hat{T}_m(\theta,\phi)$.
The total probability of state recovery is given by
\begin{equation}
q\equiv\sum_m p_{m,-m}=\sum_m
|a_{m,-\frac{1}{2}}^{(j)}(\theta,\phi)\,
 a_{m,\frac{1}{2}}^{(j)}(\theta,\phi)|^2.
\label{eq:recprob}
\end{equation}
This is the overlap between
the binomial distributions
$|a_{m,1/2}^{(j)}(\theta,\phi)|^2$ and
$|a_{m,-1/2}^{(j)}(\theta,\phi)|^2$
(see Fig.~\ref{fig1}).
The measured state can be
recovered with high probability
when these distributions overlap closely,
although the case of complete overlap
does not satisfy the condition  (\ref{eq:meacond}).
Note that when recovery occurs,
we cannot obtain any information about
the original state (\ref{eq:psi})
from the measurement outcomes $m$ and $-m$,
since the joint probability $p_{m,-m}$ does not
depend on $c_\sigma$.

If $m'\neq -m$,
we can still expect that the original state
is almost recovered as long as $m'$ is close to $-m$.
The extent to which the state of the system
is recovered can be
evaluated in terms of the fidelity between
the original state (\ref{eq:psi}) and
the final state (\ref{eq:psimm}),
\begin{align}
F_{mm'} &=
\bigl|{}_q\langle\psi |\psi_{mm'} \rangle_q\bigr| \notag \\
 &= \frac{1}{\sqrt{p_{mm'}}}
\left|\sum_{\sigma=\pm\frac{1}{2}}
a_{-m',-\sigma}^{(j)}(\theta,\phi)\,
 a_{m\sigma}^{(j)}(\theta,\phi)\,
|c_\sigma|^2\right| \notag \\
 &= \Bigl( |c_{\frac{1}{2}}|^4
\left[e_+(\theta,\phi)\right]^{m'+m}
+|c_{-\frac{1}{2}}|^4\left[e_-(\theta,\phi)
\right]^{m'+m}  \notag \\
& \quad +2|c_{\frac{1}{2}}|^2|c_{-\frac{1}{2}}|^2
\left[e_+(\theta,\phi)e_-(\theta,\phi)\right]^{\frac{m'+m}{2}}
\cos \left[(m'+m)f(\theta,\phi)\right]\Bigr)^{1/2} \notag \\
& \qquad\quad \times\Bigl(|c_{\frac{1}{2}}|^2
\left[e_+(\theta,\phi)\right]^{m'+m}
+|c_{-\frac{1}{2}}|^2
\left[e_-(\theta,\phi)\right]^{m'+m}\Bigr)^{-1/2},
\label{eq:fmm}
\end{align}
where
\begin{align}
e_\pm (\theta,\phi) &\equiv
\left(\frac{1-\chi_{\pm\frac{1}{2}}(\theta,\phi)}
{1+\chi_{\pm\frac{1}{2}}(\theta,\phi)}\right), \\
f(\theta,\phi) &\equiv
\arg\bigl[1-\sin^2\theta(\cos^2\phi+\sin^2 g) \notag \\
& \qquad\qquad\qquad+i\sin2\theta\cos \phi\sin g\bigr],
\end{align}
and $\arg[\cdots]$ represents the argument
of the complex number in the square brackets $(-\pi,\pi]$.
By definition, we obtain $F_{m,-m}=1$ as a result of
the recovery (\ref{eq:revpsi}).
It is interesting that the fidelity $F_{mm'}$ depends
on $m'+m$ but not on $j$ or on $m'-m$.
Expanding the fidelity $F_{mm'}$ 
to the second order in $m'+m$, we obtain
\begin{align}
 F_{mm'} &\sim 1- 
\frac{1}{5}|c_{\frac{1}{2}}|^2|c_{-\frac{1}{2}}|^2
\left(\frac{m'+m}{\delta m(\theta,\phi)}\right)^2 \notag \\
   &\ge  1-\frac{1}{20}
\left(\frac{m'+m}{\delta m(\theta,\phi)}\right)^2,
\end{align}
where $\delta m(\theta,\phi)$ is defined by
\begin{equation}
 \delta m(\theta,\phi) \equiv
 \sqrt{\frac{8}{5}}
\left[ \left(\ln\frac{e_+ (\theta,\phi)}
{e_- (\theta,\phi)}\right)^2+4f(\theta,\phi)^2
\right]^{-\frac{1}{2}}.
\label{eq:width}
\end{equation}
The equality is satisfied when
$|c_{1/2}|^2=|c_{-1/2}|^2=1/2$.
If the outcomes $m$ and $m'$ satisfy
\begin{equation}
  |m'+m| \le \delta m(\theta,\phi),
\label{eq:mrange}
\end{equation}
the fidelity is greater than $0.95$.
In this case,
we can say that more than 95\% of the information about
the measured state is recovered.
The total probability of this approximate recovery
is defined by
\begin{equation}
q'=\sum_{\substack{m,m' \\ F_{mm'}\ge0.95 }}p_{mm'},
\label{eq:recprob2}
\end{equation}
which depends weakly on $c_\sigma$.

\begin{figure}
\begin{center}
\includegraphics[scale=0.75]{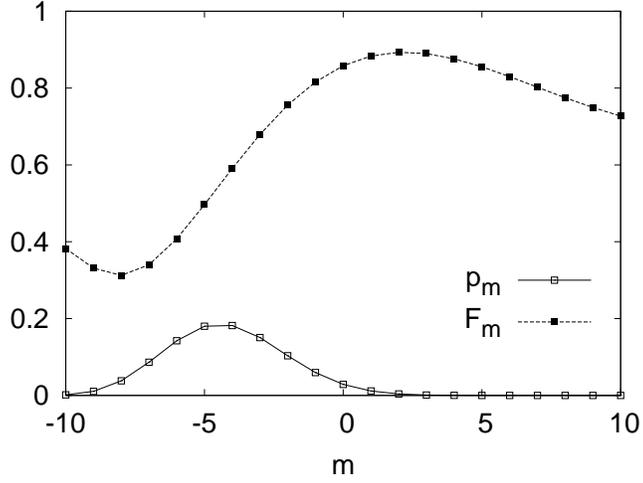}
\end{center}
\caption{\label{fig3}Probability $p_m$ and fidelity $F_m$
of the first measurement as functions of the outcome $m$
($|c_{1/2}|^2=|c_{-1/2}|^2=1/2$, $j=10$, $g=0.25$,
$\theta=\pi/6$, $\phi=\pi/6$).}
\end{figure}

\begin{figure}
\begin{center}
\includegraphics[scale=0.75]{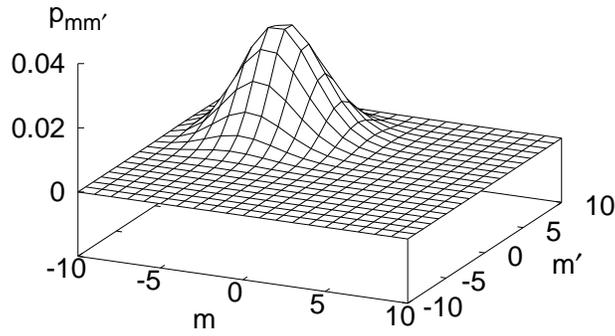}
\end{center}
\caption{\label{fig4}Joint probability $p_{mm'}$
of the first and second measurements
as a function of the outcomes $m$ and $m'$
($|c_{1/2}|^2=|c_{-1/2}|^2=1/2$, $j=10$, $g=0.25$,
$\theta=\pi/6$, $\phi=\pi/6$).}
\end{figure}

\begin{figure}
\begin{center}
\includegraphics[scale=0.75]{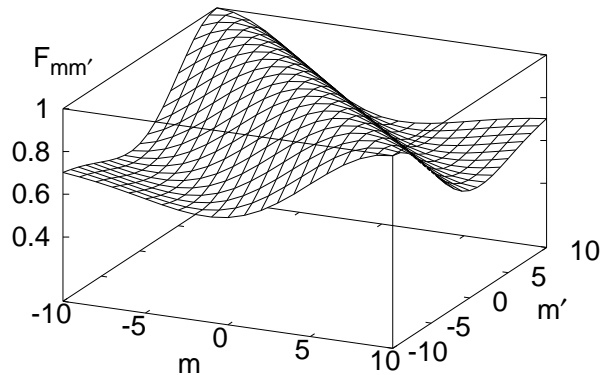}
\end{center}
\caption{\label{fig5}Fidelity $F_{mm'}$
after the second measurement
as a function of the outcomes $m$ and $m'$
($|c_{1/2}|^2=|c_{-1/2}|^2=1/2$, $j=10$, $g=0.25$,
$\theta=\pi/6$, $\phi=\pi/6$).
$F_{mm'}$ depends only on $m'+m$ with $F_{m,-m}=1$.}
\end{figure}

As an example, we consider the case where
$|c_{1/2}|^2=|c_{-1/2}|^2=1/2$,
$j=10$, $g=0.25$, $\theta=\pi/6$, and $\phi=\pi/6$.
This is the worst case for which
the lower bound in Eq.~(\ref{eq:width}) is achieved.
Figure \ref{fig3} shows
the probability (\ref{eq:pm}) and the fidelity (\ref{eq:fm})
of the first measurement $\{\hat{T}_m(\pi/6,\pi/6)\}$
as functions of the outcome $m$.
The average fidelity after the first measurement
is $\sum_m p_m F_m=0.57$.
To recover the fidelity lost by the first measurement,
the second measurement $\{\hat{T}_m(5\pi/6,5\pi/6)\}$
is performed.
Figure \ref{fig4} shows
the probability (\ref{eq:pmm})
as a function of the outcomes $m$ for
the first measurement and $m'$ for
the second measurement.
The two peaks (\ref{eq:twopeak})
of the joint probability merge into a single peak
located on the line of recovery ($m'=-m$),
since $\chi_{1/2}(\theta,\phi)$
and $\chi_{-1/2}(\theta,\phi)$
are close to each other.
This indicates that the highly probable events
are concentrated near the line of recovery.
In fact,
the total probability of recovery (\ref{eq:recprob})
becomes large due to the large overlap of
$|a_{m,1/2}^{(j)}(\theta,\phi)|^2$ and
$|a_{m,-1/2}^{(j)}(\theta,\phi)|^2$.
In this example, we obtain $q=0.13$.
The more tolerable is the error in terms of the fidelity,
the larger is the probability of recovery.
Figure \ref{fig5} shows
the fidelity (\ref{eq:fmm})
after the second measurement
as a function of the outcomes $m$ and $m'$.
The average fidelity after the second measurement
is $\sum_{mm'} p_{mm'} F_{mm'}=0.93$.
The fidelity is larger than $0.95$ provided that
$|m'+m|$ is less than $\delta m(\theta,\phi)=2.3$
defined by Eq.~(\ref{eq:width}).
The total probability of approximate recovery
(\ref{eq:recprob2}) is $q'=0.57$.

\subsection{Information Gain versus Fidelity Loss}
As noted in the preceding subsection,
we cannot obtain any information about the measured state
if a successful recovery occurs by a reversing measurement.
In other words, successful recovery obliterates
the information obtained by the first measurement.
Therefore, one might think that
it is not worthwhile performing a reversing measurement.
However, when the recovery is only partially successful,
the reversing measurement can
improve the fidelity together with providing further information.
We show this here by a simple situation.

\begin{figure}
\begin{center}
\includegraphics[scale=0.695]{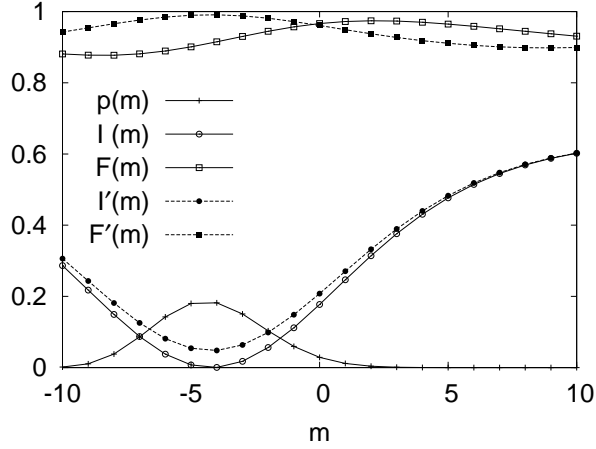}
\end{center}
\caption{\label{fig6}Probability $p(m)$ of
obtaining outcome $m$ for the first measurement, and
the corresponding information gain $I(m)$ and fidelity $F(m)$,
with $j=10$, $g=0.25$, $\theta=\pi/6$, $\phi=\pi/6$,
and $\gamma=\pi/6$.
Also shown are the expected information gain $I'(m)$ and
expected fidelity $F'(m)$
after the reversing measurement,
given that the outcome of the first measurement is $m$.}
\end{figure}

Suppose that the state of the system is known to be
either $\ket{a}_q$ or $\ket{b}_q$
with equal probability, $p(a)=p(b)=1/2$, where
we choose the two states as
\begin{align}
  \ket{a}_q &= \cos\frac{\gamma}{2}\,\ket{1/2}_q
     +\sin\frac{\gamma}{2}\,\ket{-1/2}_q, \\
  \ket{b}_q &= -\sin\frac{\gamma}{2}\,\ket{1/2}_q
     +\cos\frac{\gamma}{2}\,\ket{-1/2}_q,
\end{align}
with $\gamma$ being a real constant ($0<\gamma<\pi/2$).
The Shannon entropy associated with the system
is initially given by
\begin{equation}
  H_0=-p(a)\log_2 p(a)-p(b)\log_2 p(b)=1,
\end{equation}
which is a measure of the lack of information about the system.
We then perform the measurement $\{\hat{T}_m(\theta,\phi)\}$
in an attempt to obtain information about the system.
If the input state of the system is $\ket{a}_q$,
the measurement yields an outcome $m$
with probability $p(m|a)$,
and the postmeasurement state is given by $\ket{a_m}_q$
whose fidelity to $\ket{a}_q$ is $F(m,a)$.
Here the probability $p(m|a)$,
the postmeasurement state $\ket{a_m}_q$, and
the fidelity $F(m,a)$ can be evaluated
according to Eqs.~(\ref{eq:pm}),
(\ref{eq:psim}), and (\ref{eq:fm}).
Similarly, if the input state of the system is $\ket{b}_q$,
the corresponding probability, the postmeasurement state,
and the fidelity are given by $p(m|b)$, $\ket{b_m}_q$,
and $F(m,b)$, respectively.
The total probability for outcome $m$ is
$p(m)=p(m|a)\,p(a)+p(m|b)\,p(b)$.
Suppose that we obtain the outcome $m$.
Then Bayes' rule tells us that the probability that
the input state is $\ket{a}_q$ (or $\ket{b}_q$)
is given by $p(a|m)=p(m|a)\,p(a)/p(m)$
[or $p(b|m)=p(m|b)\,p(b)/p(m)$].
The Shannon entropy after the measurement
with outcome $m$ becomes
\begin{equation}
  H(m)=-p(a|m)\log_2 p(a|m)-p(b|m)\log_2 p(b|m).
\end{equation}
This means that the amount of information
obtained from the outcome $m$ is
\begin{equation}
 I(m)=H_0-H(m).
\end{equation}
The average fidelity for a given outcome $m$ is given by
\begin{equation}
 F(m)=F(m,a)\,p(a|m)+F(m,b)\,p(b|m).
\end{equation}
Figure \ref{fig6} shows
the probability for outcome $p(m)$,
the information gain $I(m)$, and the fidelity $F(m)$
as functions of $m$
for $j=10$, $g=0.25$, $\theta=\pi/6$, $\phi=\pi/6$,
and $\gamma=\pi/6$.
We find that
an outcome that is realized with a high probability gives
less information than one with a low probability.

After obtaining the outcome $m$ for
the measurement $\{\hat{T}_m(\theta,\phi)\}$,
we perform the reversing measurement
$\{\hat{T}_m(\pi-\theta,\pi-\phi)\}$ to
recover the measured state.
Let $m'$ be the outcome of the reversing measurement.
If the input state of the system is $\ket{a}_q$
before the first measurement $\{\hat{T}_m(\theta,\phi)\}$,
the joint probability for a pair of outcomes $(m,m')$
is given by $p(m,m'|a)$ and
the corresponding postmeasurement state is $\ket{a_{mm'}}_q$
whose fidelity to $\ket{a}_q$ is $F(m,m',a)$.
We can calculate the probability $p(m,m'|a)$,
the postmeasurement state $\ket{a_{mm'}}_q$, and
the fidelity $F(m,m',a)$
according to Eqs.~(\ref{eq:pmm}),
(\ref{eq:psimm}), and (\ref{eq:fmm}).
Similarly, if the input state of the system is $\ket{b}_q$,
the joint probability, the postmeasurement state,
and the fidelity can be calculated to give
$p(m,m'|b)$, $\ket{b_{mm'}}_q$, and $F(m,m',b)$, respectively.
The total joint probability for a pair of outcomes $(m,m')$ is
$p(m,m')=p(m,m'|a)\,p(a)+p(m,m'|b)\,p(b)$.
From the two outcomes $(m,m')$,
we know that the input state is $\ket{a}_q$
with probability $p(a|m,m')=p(m,m'|a)\,p(a)/p(m,m')$
and is $\ket{b}_q$
with probability $p(b|m,m')=p(m,m'|b)\,p(b) / p(m,m')$.
The Shannon entropy after the two measurements
with outcomes $(m,m')$ becomes
\begin{equation}
  H(m,m')=-p(a|m,m')\log_2 p(a|m,m')
            -p(b|m,m')\log_2 p(b|m,m').
\end{equation}
The amount of obtained information is given by
\begin{equation}
  I(m,m')=H_0-H(m,m'),
\end{equation}
and the fidelity becomes
\begin{equation}
  F(m,m')=F(m,m',a)\,p(a|m,m')+F(m,m',b)\,p(b|m,m').
\end{equation}

When the two outcomes satisfy $m'=-m$, recovery is achieved by
the reversing measurement, $F(m,-m,a)=F(m,-m,b)=1$.
We cannot then obtain any information about the system
because $p(m,-m|a)=p(m,-m|b)$, i.e., $I(m,-m)=0$
[note, however, that $I(m)>0$].
However, if $m'\sim-m$, we might expect a partial recovery
should be achieved with some information loss.
To check this,
we consider the expectation value of the information
to be obtained by performing the reversing measurement,
given the outcome $m$ of the first measurement
with information $I(m)$.
Since the conditional probability of obtaining outcome $m'$
for the reversing measurement is $p(m'|m)=p(m,m')/p(m)$,
the expectation value of the information is given by
\begin{equation}
 I'(m)=\sum_{m'} p(m'|m)\,I(m,m'),
\end{equation}
while the expectation value of the fidelity is given by
\begin{equation}
 F'(m)=\sum_{m'} p(m'|m)\,F(m,m').
\end{equation}
The expectation value of the information gain $I'(m)$
and that of the fidelity $F'(m)$ are
shown in Fig.~\ref{fig6} as functions of $m$
for $j=10$, $g=0.25$, $\theta=\pi/6$, $\phi=\pi/6$,
and $\gamma=\pi/6$.
Note that $F'(m)>F(m)$ and $I'(m)>I(m)$ for several outcomes.
This implies that the reversing measurement
can achieve both a partial recovery of the quantum state
and \emph{further information gain}
rather than information loss.

\subsection{Effect of Probe Spin}
\begin{figure}
\begin{center}
\includegraphics[scale=0.75]{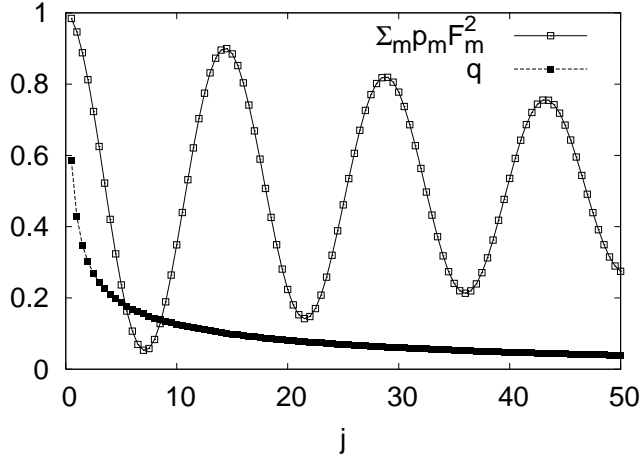}
\end{center}
\caption{\label{fig7}Average squared fidelity
after the first measurement $\sum_m p_m F_m^2$ and
total probability of recovery $q$ as functions of $j$
($|c_{1/2}|^2=|c_{-1/2}|^2=1/2$, $j=10$, $g=0.25$,
$\theta=\pi/6$, $\phi=\pi/6$).
Although the strength of the interaction $g$
is much smaller ($\sim10^{-8}$)
in the real situation discussed in Sec.~\ref{sec:exp},
it can be enhanced by a cavity-assisted interaction or
by collective enhancement via large $j$ and $s$.}
\end{figure}

We discuss here the effect of a high-spin probe ($j>1/2$).
In this case, the recovery of the measured state
emerges more clearly
because of the large change in the fidelity
during the measurements.
To simplify the calculations, we consider here
the average squared fidelity after the first measurement,
given by
\begin{equation}
\sum_m p_m F_m^2
= |c_{\frac{1}{2}}|^4+|c_{-\frac{1}{2}}|^4 
+2|c_{\frac{1}{2}}|^2|c_{-\frac{1}{2}}|^2
\,h(\theta)^j \cos\left[jk(\theta)\right]
\label{eq:asfidelity}
\end{equation}
with
\begin{align}
 h(\theta) &\equiv 1-\sin^2\theta\sin^2g, \\
 k(\theta) &\equiv 2\arg\bigl[\cos g-i\sin g\cos\theta\bigr].
\end{align}
Figure \ref{fig7} shows the average squared fidelity
as a function of $j$,
exhibiting a damped oscillation with period $2\pi/|k(\theta)|$.
The oscillation results from
$\arg [a_{m\sigma}^{(j)}(\theta,\phi)]$,
which changes the relative phase
between the states $\ket{1/2}_q$ and $\ket{-1/2}_q$.
When the probe has a high spin ($j>1/2$),
a large fidelity can be lost
as a result of the first measurement.
In particular,
the fidelity loss becomes maximal at $j\sim\pi/|k(\theta)|$.
Nevertheless, such a large fidelity loss
can be recovered as a result of the second measurement,
as discussed in the preceding subsection.

Of course, as a tradeoff,
the total probability of recovery (\ref{eq:recprob})
becomes small, as shown in Fig.~\ref{fig7}.
For large $j$, the central limit theorem (\ref{eq:clt})
gives an exponential decay of the probability of recovery,
\begin{equation}
q  \sim
\frac{1}{\sqrt{2\pi j v(\theta,\phi)}}
\exp\left[-\frac{j(\chi_{\frac{1}{2}}(\theta,\phi)
-\chi_{-\frac{1}{2}}(\theta,\phi))^2}%
{2v(\theta,\phi)}\right],
\end{equation}
where
\begin{equation}
 v(\theta,\phi)\equiv
1-\frac{1}{2}
\left(\chi_{\frac{1}{2}}(\theta,\phi)^2
+\chi_{-\frac{1}{2}}(\theta,\phi)^2\right).
\end{equation}
This decay results from the fact that
as $j$ increases,
the two peaks (\ref{eq:twopeak}) of the joint probability
split away from the line of recovery ($m'=-m$) and
therefore the probability on the line decreases exponentially.
Similarly, the total probability
of approximate recovery (\ref{eq:recprob2}) also
decreases exponentially as $j$ increases,
since the increase of $j$ cannot expand
the width (\ref{eq:width}) for approximate recovery.
Due to the decrease in the probability of recovery,
the average squared fidelity after the second measurement
also decreases as
\begin{equation}
\sum_{m,m'} p_{mm'} F_{mm'}^2
= |c_{\frac{1}{2}}|^4+|c_{-\frac{1}{2}}|^4 
+2|c_{\frac{1}{2}}|^2|c_{-\frac{1}{2}}|^2\,h(\theta)^{2j}.
\end{equation}
This fidelity does not oscillate,
unlike the case in Eq.~(\ref{eq:asfidelity}),
because the change in the relative phase during the first measurement
is on average canceled by that during the second measurement.

\subsection{Quantum Fluctuation of Probe Spin}
So far, the spin $j$ of the probe
has been assumed to be a definite value.
However, some physical systems
are described by indefinite spin.
For example, a two-mode laser
is regarded as a spin system with indefinite spin
because of quantum fluctuations in the number of photons
(see Sec.~\ref{sec:exp}).
We here show that even when
the spin of the probe is affected by quantum fluctuations,
the measurement $\{\hat{T}_m(\pi-\theta,\pi-\phi)\}$
remains a reversing measurement of
the measurement $\{\hat{T}_m(\theta,\phi)\}$.

When the probe spin $j$ fluctuates
quantum-mechanically,
the initial probe state (\ref{eq:probinit})
is replaced with
\begin{equation}
\ket{\theta,\phi}_p=
\exp\left(-\frac{i}{\hbar}\hat{J}_{z}\phi\right)
\exp\left(-\frac{i}{\hbar}\hat{J}_{y}\theta\right)
\sum_j b_j\ket{j}_p,
\end{equation}
where $j=0,1/2,1,3/2,\ldots$ and
the coefficients $\{b_j\}$ satisfy
the normalization condition $\sum_j |b_j|^2=1$.
Note that a measurement yielding an outcome $m$
($=0,\pm1/2,\pm1,\pm3/2,\ldots$)
eliminates probe states with $j\neq|m|,|m|+1,|m|+2,\ldots$,
since
\begin{equation}
  \sum_j\sum_{m=-j}^j=\sum_m \sum_{j\ge |m|}',
\end{equation}
where the prime indicates summation over $j$
such that $j-|m|$ is a nonnegative integer.
The measurement operators (\ref{eq:halfspinop})
and (\ref{eq:halfspinop2}) are thus replaced with
\begin{equation}
{}_q\bra{\sigma'}\, \hat{T}_m(\theta,\phi)\,\ket{\sigma}_q
  = \sum_{j\ge |m|}' b_j\left(\begin{array}{cc}
    a_{m,\frac{1}{2}}^{(j)}(\theta,\phi) & 0 \\
    0 & a_{m,-\frac{1}{2}}^{(j)}(\theta,\phi)
    \end{array}\right),
\end{equation}
and
\begin{align}
& {}_q\bra{\sigma'}\, \hat{T}_{m'}(\pi-\theta,\pi-\phi)\,
  \ket{\sigma}_q \notag \\
&\qquad\quad= e^{-im'\pi}\sum_{j\ge |m'|}' b_j
    \left(\begin{array}{cc}
    a_{-m',-\frac{1}{2}}^{(j)}(\theta,\phi) & 0 \\
    0 & a_{-m',\frac{1}{2}}^{(j)}(\theta,\phi)
    \end{array}\right),
\end{align}
respectively.
It is easy to see that
$\hat{T}_{-m}(\pi-\theta,\pi-\phi)\,\hat{T}_m(\theta,\phi)$
is proportional to the identity operator,
\begin{align}
& \hat{T}_{-m}(\pi-\theta,\pi-\phi)\,\hat{T}_m(\theta,\phi) \notag \\
&\qquad\quad = \left[e^{im\pi}
  \sum_{j\ge |m|}' \sum_{j'\ge |m|}' b_j\, b_{j'}\,
  a_{m,-\frac{1}{2}}^{(j)}(\theta,\phi)\,
 a_{m,\frac{1}{2}}^{(j')}(\theta,\phi)\right]\hat{I}.
\end{align}
Consequently,
the measurement $\{\hat{T}_m(\pi-\theta,\pi-\phi)\}$
is still a reversing measurement of
the measurement $\{\hat{T}_m(\theta,\phi)\}$
in the presence of quantum fluctuations.
In contrast,
the measurement $\{\hat{T}_m(\pi-\theta,\pi-\phi)\}$
is no longer a reversing measurement of
the measurement $\{\hat{T}_m(\theta,\phi)\}$
if the probe spin is affected by classical fluctuations
that replace the probe state (\ref{eq:probinit})
with a mixed state.

This tolerance for quantum fluctuation of the probe spin
is emphasized when we consider
the measurement $\{\hat{T}_m(\pi-\theta,-\phi)\}$.
This is another reversing measurement of
the measurement $\{\hat{T}_m(\theta,\phi)\}$,
since
\begin{equation}
 a_{m'\sigma}^{(j)}(\pi-\theta,-\phi)=
 (-1)^{j+m'}
  a_{m',-\sigma}^{(j)}(\theta,\phi)
\label{eq:asymm2}
\end{equation}
holds, rather than Eq.~(\ref{eq:asymm}).
The measured state is recovered
if the outcome of the second measurement is
the same as that of the first ($m'=m$).
As long as the spin $j$ of the probe
has a definite value,
this reversing measurement is equivalent to
the measurement $\{\hat{T}_m(\pi-\theta,\pi-\phi)\}$.
However, when the probe spin is affected
by quantum fluctuation,
the measurement $\{\hat{T}_m(\pi-\theta,-\phi)\}$
is no longer a reversing measurement of
the measurement $\{\hat{T}_m(\theta,\phi)\}$
due to the $j$-dependent factor $(-1)^{j+m'}$
in Eq.~(\ref{eq:asymm2}).

\section{\label{sec:rev-mea2}%
Reversing Measurement on a High-spin \\
System ($s>1/2$)}
We next consider a reversing measurement of
a physically reversible measurement
$\{\hat{T}_m(\theta,\phi)\}$ for the case of
measured systems with $s>1/2$.
Provided that the condition (\ref{eq:revcond}) is satisfied,
the physical reversibility implies the existence of
a reversing measurement~\cite{UeImNa96,Ueda97}.
More specifically,
for a first measurement with outcome $m$,
we consider a second measurement
$\{\hat{R}^{(m)}_0, \hat{R}^{(m)}_1\}$
with two possible outcomes, say $0$ and $1$, such that
\begin{align}
   \hat{R}^{(m)}_0 &=\kappa_m \sum^s_{\sigma=-s}
  \left[ a_{m\sigma}^{(j)}(\theta,\phi)\right]^{-1}
\,\ket{\sigma}_q {}_q\bra{\sigma}, \\
\hat{R}^{(m)}_1 &=
\sqrt{\hat{I}-\hat{R}^{(m)\dagger}_0\hat{R}^{(m)}_0},
\end{align}
where $\kappa_m$ is a nonzero constant.
If this measurement yields the outcome $0$,
the original state of the system is restored because
\begin{equation}
  \hat{R}^{(m)}_0\, \hat{T}_m(\theta,\phi) =\kappa_m \hat{I},
\end{equation}
as seen from Eq.~(\ref{eq:meaop}).
Unfortunately, the physical implementation of
this measurement is not obvious.
Instead,
we consider an approximate reversing measurement
that has a clear physical implementation using
the measurement $\{\hat{T}_m(\pi-\theta,\pi-\phi)\}$.
Unlike the case of $s=1/2$,
the measurement $\{\hat{T}_m(\pi-\theta,\pi-\phi)\}$
is not an exact reversing measurement,
since $\hat{T}_{-m}(\pi-\theta,\pi-\phi)$ is not proportional to
the inverse of $\hat{T}_m(\theta,\phi)$.
Contrary to Eq.~(\ref{eq:proto}), we have
\begin{equation}
\hat{T}_{-m}(\pi-\theta,\pi-\phi)\,\hat{T}_m(\theta,\phi)
\not\propto \hat{I}.
\end{equation}
Nevertheless,
there are two physical situations in which
the measurement $\{\hat{T}_m(\pi-\theta,\pi-\phi)\}$
serves approximately as a reversing measurement for
the original measurement $\{\hat{T}_m(\theta,\phi)\}$:
(i) the measured state can be confined to
a two-dimensional subspace or
(ii) the interaction
between the system and probe is sufficiently weak.
In this section, we describe these approximate
schemes for the reversing measurement.

\subsection{Two-dimensional Subspace Model}
We assume that
the initial state of the measured system with spin $s$
is in a two-dimensional subspace spanned by
$\{\ket{\tilde{\sigma}},\ket{-\tilde{\sigma}}\}$,
where $\tilde{\sigma}\hbar$ is any one of
the nonzero eigenvalues of $\hat{S}_{z}$.
That is, we know \textit{a priori} that
\begin{equation}
   \ket{\psi}_q=\sum_{\sigma=\pm\tilde{\sigma}}
c_\sigma \ket{\sigma}_q,
\label{eq:twodim}
\end{equation}
instead of the general state (\ref{eq:psi}).
Since the measurement operator
is diagonal, as in Eq.~(\ref{eq:meaop}),
the state of the system remains in this subspace
after the measurement.
The measurement operators $\hat{T}_m(\theta,\phi)$
and $\hat{T}_{m'}(\pi-\theta,\pi-\phi)$ are thus represented by
$2\times2$ matrices within this subspace.
These matrices are identical to those in the $s=1/2$ case
[see Eqs.~(\ref{eq:halfspinop}) and (\ref{eq:halfspinop2})]
with the strength of the interaction given by
\begin{equation}
  g'= 2g\tilde{\sigma}.
\label{eq:renint}
\end{equation}
Consequently, the measurement $\{\hat{T}_m(\pi-\theta,\pi-\phi)\}$
is a reversing measurement of the measurement $\{\hat{T}_m(\theta,\phi)\}$
when the initial state of the system
is confined to the two-dimensional subspace.

\begin{figure}
\begin{center}
\includegraphics[scale=0.75]{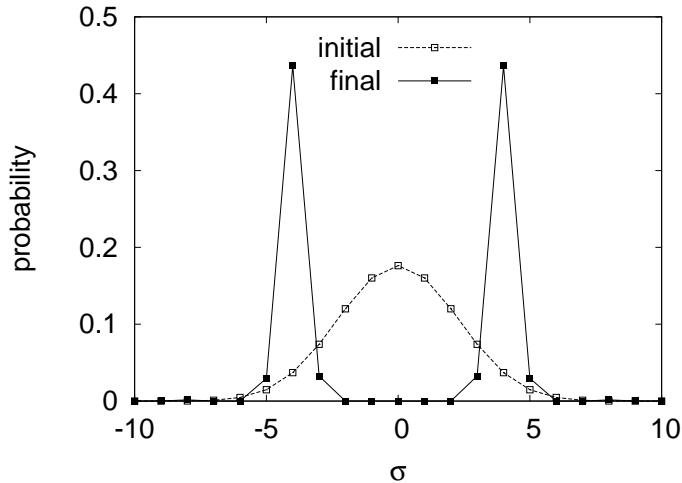}
\end{center}
\caption{\label{fig8}Initial spin distribution
$|c'_\sigma|^2$ and final spin distribution $\rho_m(\sigma)$
as functions of $\sigma$
($j=s=10$, $g=0.25$, $m=5$).
$\rho_m(\sigma)$
has a pair of highest peaks at $\sigma=\pm4$
(the other peaks are too small to be seen on the scale
of this figure).
The probability $p'_m$ in Eq.~(\ref{eq:twodimprob})
is calculated to be $0.016$.}
\end{figure}

The analysis of fidelity in this model
is the same as that in the case where $s=1/2$
in the preceding section,
provided that the renormalized
strength of the interaction (\ref{eq:renint}) is used.
The remaining problem is preparing the system
in the two-dimensional subspace.
In order to prepare the state (\ref{eq:twodim}),
we here use the scheme in Ref.~\cite{THTTIY99},
which was originally proposed
to realize a squeezed spin state~\cite{KitUed93}.
The system is first prepared in the state
\begin{align}
\ket{\psi'}_q &=
\exp\left(-\frac{i}{\hbar}\hat{S}_{z}\varphi\right)
\exp\left(-\frac{i}{\hbar}\hat{S}_{y}\frac{\pi}{2}\right)\ket{s}_q
\notag \\
 &= \sum_\sigma 
 c'_\sigma \,\ket{\sigma}_q,
\end{align}
where
\begin{equation}
 c'_\sigma\equiv e^{-i\sigma\varphi}\,
d_{\sigma s}^{(s)}\left(\frac{\pi}{2}\right)
=e^{-i\sigma\varphi}\,
\frac{1}{2^{s}}\sqrt{\frac{(2s)!}{(s+\sigma)!(s-\sigma)!}}.
\end{equation}
This is a coherent spin state~\cite{ACGT72} and
is an eigenstate of the spin component
$\hat{S}_{\varphi}=\hat{S}_{x}\cos\varphi+\hat{S}_{y}\sin\varphi$
with eigenvalue $s\hbar$.
Performing the measurement $\{\hat{T}_m(\pi/2,0)\}$ on this state
 yields an outcome $m$ with probability
\begin{equation}
 p'_{m} = \sum_{\sigma}
\left|a_{m\sigma}^{(j)}\left(\frac{\pi}{2},0\right)\right|^2
|c'_\sigma|^2
\label{eq:twodimprob}
\end{equation}
and then causes state reduction to
\begin{equation}
\ket{\psi'_m}_q=\frac{1}{\sqrt{p'_m}}
\sum_{\sigma} a_{m\sigma}^{(j)}\left(\frac{\pi}{2},0\right)\,
c'_\sigma \ket{\sigma}_q.
\label{eq:psipm}
\end{equation}
The spin distribution of this state
$\rho_m(\sigma)\equiv |{}_q\langle\sigma
|\psi'_m \rangle_q|^2$ is given by
\begin{align}
 \rho_m(\sigma) &=
    \frac{1}{p'_m}
   \left|a_{m\sigma}^{(j)}\left(\frac{\pi}{2},0\right)\right|^2
   |c'_\sigma|^2 \notag \\
  &= \left[ \frac{1}{p'_m}\frac{(2j)!}{(j+m)!(j-m)!} \right]
     \left[\frac{1}{2^{2s}}
           \frac{(2s)!}{(s+\sigma)!(s-\sigma)!}\right] \notag \\
  &  \qquad\qquad{}\times  \left[\cos^2(g\sigma)\right]^{j-m}
   \left[\sin^2(g\sigma)\right]^{j+m}. \label{eq:sd}
\end{align}
Clearly, this distribution satisfies
$\rho_m(0)=0$ (if $j\neq -m$) and
$\rho_m(\sigma)=\rho_m(-\sigma)$,
and is damped by the second binomial factor
for large $|\sigma|$.
These facts imply that when $j\neq -m$,
the spin distribution has a pair of highest peaks
at $\sigma=\pm\tilde{\sigma}_m$ (see Fig.~\ref{fig8}),
where $\tilde{\sigma}_m$ is evaluated as
\begin{equation}
  \tilde{\sigma}_m  \sim \frac{1}{g}\arctan\sqrt{\frac{j+m}{j-m}}
\end{equation}
if $g \ll \pi/2<gs$.
The state (\ref{eq:psipm}) can thus be approximated as
\begin{equation}
\ket{\psi'_m}_q \sim \frac{1}{\sqrt{2}} \Bigl[
 e^{-i\tilde{\sigma}_m\varphi} \ket{\tilde{\sigma}_m}_q
+(-1)^{j+m} e^{i\tilde{\sigma}_m\varphi}
\ket{-\tilde{\sigma}_m}_q \Bigr],
\end{equation}
where the relative phase is determined from the identity
$a_{m,-\sigma}^{(j)}(\pi/2,0)=(-1)^{j+m}a_{m\sigma}^{(j)}(\pi/2,0)$.
According to Eq.~(\ref{eq:sd}),
this is a good approximation for large $j$.
Finally, by performing a further measurement
$\{\hat{T}_m(\theta',\phi')\}$ on this state,
we can prepare a state in the form of
\begin{equation}
\ket{\psi}_q=\sum_{\sigma=\pm\tilde{\sigma}_m}
  c_\sigma \ket{\sigma}_q,
\end{equation}
where the coefficients depend on
the angles $(\theta',\phi')$ and the outcome.

\subsection{Weak-interaction Model}
We next consider another physical situation for
the approximate reversing measurement
$\{\hat{T}_m(\pi-\theta,\pi-\phi)\}$.
We assume that the interaction is so weak that
the measurement operators can be expanded
in powers of $g$.
We then obtain
\begin{equation}
\hat{T}_{-m}(\pi-\theta,\pi-\phi)\,\hat{T}_m(\theta,\phi)
\sim \left[e^{im\pi} a_{m,0}^{(j)}(\theta,\phi)^2\right]\hat{I}
 +O(g^2).
\label{eq:proptow}
\end{equation}
This means that the measurement $\{\hat{T}_m(\pi-\theta,\pi-\phi)\}$
is a reversing measurement of
the measurement $\{\hat{T}_m(\theta,\phi)\}$
to an accuracy of the order of $g$.

As shown below, the second-order term,
which is neglected in Eq.~(\ref{eq:proptow}),
does not affect the fidelity up to the order of $g^3$.
For the two successive measurements $\{\hat{T}_m(\theta,\phi)\}$
and $\{\hat{T}_m(\pi-\theta,\pi-\phi)\}$,
we define the joint probability, the final state,
and the fidelity, as in the case of $s=1/2$, by
\begin{align}
p_{mm'} &= \sum_{\sigma}
|a_{-m',-\sigma}^{(j)}(\theta,\phi)\,
 a_{m\sigma}^{(j)}(\theta,\phi)|^2 |c_\sigma|^2, \label{eq:pmmw} \\
\ket{\psi_{mm'}}_q &=\frac{e^{-im'\pi}}{\sqrt{p_{mm'}}}
\sum_{\sigma} a_{-m',-\sigma}^{(j)}(\theta,\phi)\,
 a_{m\sigma}^{(j)}(\theta,\phi)\,c_\sigma \ket{\sigma}_q,
\end{align}
and
\begin{equation}
F_{mm'}  = \frac{1}{\sqrt{p_{mm'}}}
\left|\sum_{\sigma}
a_{-m',-\sigma}^{(j)}(\theta,\phi)\,
 a_{m\sigma}^{(j)}(\theta,\phi)\,
|c_\sigma|^2\right| , \label{eq:fmmw}
\end{equation}
respectively, using the relation (\ref{eq:asymm}).
Expanding the fidelity up to the second order in $g$, we obtain
\begin{align}
F_{mm'}  &\sim 1- \frac{1}{20}
\left[\frac{\overline{\sigma^2}-(\overline{\sigma})^2}{s^2}\right]
\left(\frac{m'+m}{\delta\widetilde{m}(\theta,\phi)}\right)^2 \notag \\
   &\ge  1-\frac{1}{20}
\left(\frac{m'+m}{\delta\widetilde{m}(\theta,\phi)}\right)^2,
\label{eq:fmmw2}
\end{align}
where
\begin{equation}
 \overline{\sigma}\equiv \sum_\sigma\sigma |c_\sigma|^2, \qquad
\overline{\sigma^2}\equiv \sum_\sigma\sigma^2 |c_\sigma|^2,
\end{equation}
\begin{equation}
  \delta \tilde{m}(\theta,\phi) \equiv
 \frac{1}{2\sqrt{10}\,s}\left(
 \frac{\sqrt{1-\sin^2\theta\cos^2\phi}}
{|g\sin\theta |}\right).
\label{eq:widthw}
\end{equation}
Consequently, we find that
when the two outcomes cancel each other ($m'=-m$),
the information about the original state
is restored to within the accuracy of $g^3$,
because $F_{m,-m}\sim 1+O(g^4)$.
The measurement $\{\hat{T}_m(\pi-\theta,\pi-\phi)\}$
is thus a reversing measurement of
the measurement $\{\hat{T}_m(\theta,\phi)\}$
if the fourth-order term in $g$ can be neglected.
Evaluating the fourth-order term, we obtain
the condition for the strength of the interaction as
\begin{equation}
 g^4 \ll \frac{1}{s^4j^2}
\left(\frac{1-|\sin\theta\cos\phi|}{\sqrt{2}\sin\theta}\right)^2.
\label{eq:weakcond}
\end{equation}
As in the case of $s=1/2$,
we define the total probability of approximate recovery by
\begin{equation}
q'=\sum_{\substack{m,m' \\ F_{mm'}\ge0.95 }}p_{mm'},
\label{eq:recprobw}
\end{equation}
where a sufficient condition for $F_{mm'}\ge0.95$ is given by
\begin{equation}
  |m'+m| \le \delta\widetilde{m}(\theta,\phi).
\end{equation}

\begin{figure}
\begin{center}
\includegraphics[scale=0.75]{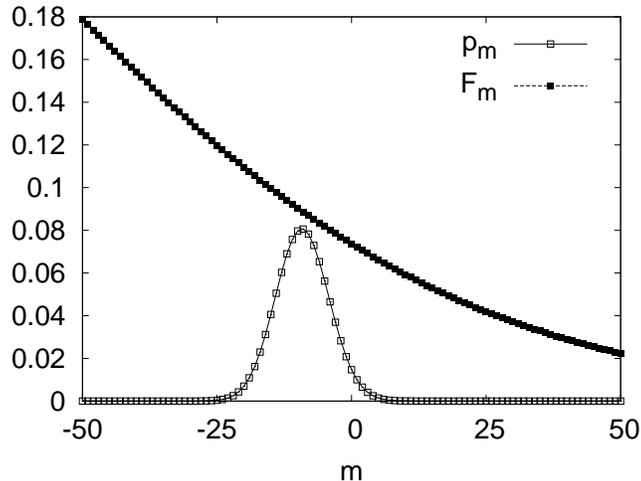}
\end{center}
\caption{\label{fig9}Probability $p_m$ and fidelity $F_m$
of the first measurement on the state $\ket{S_{x}=s\hbar}_q$
as functions of the outcome $m$
($j=50$, $s=10$, $g=0.01$, $\theta=\pi/12$, $\phi=\pi/4$).}
\end{figure}

\begin{figure}
\begin{center}
\includegraphics[scale=0.75]{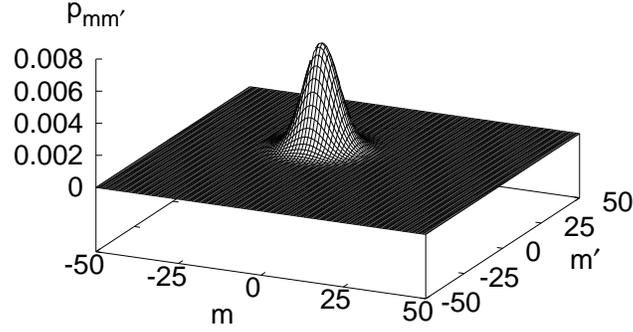}
\end{center}
\caption{\label{fig10}Joint probability $p_{mm'}$
of the first and second measurements
on the state $\ket{S_{x}=s\hbar}_q$
as a function of the outcomes $m$ and $m'$
($j=50$, $s=10$, $g=0.01$, $\theta=\pi/12$, $\phi=\pi/4$).}
\end{figure}

\begin{figure}
\begin{center}
\includegraphics[scale=0.75]{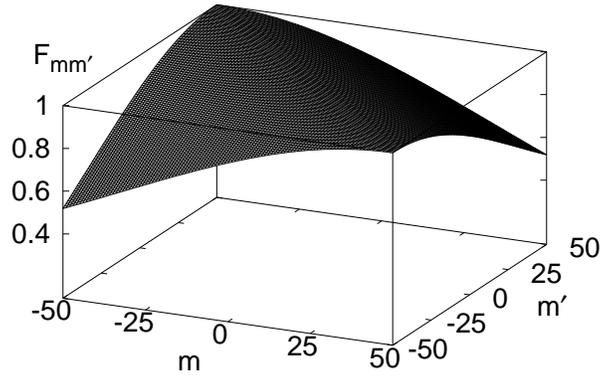}
\end{center}
\caption{\label{fig11}Fidelity $F_{mm'}$
after the two measurements on the state $\ket{S_{x}=s\hbar}_q$
as a function of the outcomes $m$ and $m'$
($j=50$, $s=10$, $g=0.01$, $\theta=\pi/12$, $\phi=\pi/4$).}
\end{figure}

As an example, we consider the case where
$j=50$, $s=10$, $g=0.01$, $\theta=\pi/12$, and $\phi=\pi/4$.
The measured system is assumed to be
in a coherent spin state
\begin{equation}
\ket{\psi}_q=\ket{S_{x}=s\hbar}_q =
\exp\left(-\frac{i}{\hbar}\hat{S}_{y}\frac{\pi}{2}\right)
\ket{s}_q,
\label{eq:css}
\end{equation}
which is the eigenstate of $\hat{S}_{x}$ with eigenvalue $s\hbar$.
Figure \ref{fig9} shows
the probability (\ref{eq:pm}) and the fidelity (\ref{eq:fm})
of the first measurement $\{\hat{T}_m(\pi/12,\pi/4)\}$
as functions of the outcome $m$.
The average fidelity after the first measurement
is $\sum_m p_m F_m=0.089$.
The second measurement $\{\hat{T}_m(11\pi/12,3\pi/4)\}$
is then performed.
Figure \ref{fig10} shows
the probability (\ref{eq:pmmw})
as a function of the outcomes $m$ for
the first measurement and $m'$ for
the second measurement.
Figure \ref{fig11} shows
the fidelity (\ref{eq:fmmw})
after the second measurement
as a function of the outcomes $m$ and $m'$.
Although the fidelity $F_{mm'}$ may depend on $j$ and on $m'-m$
if $s>1/2$, it approximately depends only
on $m'+m$ [see Eq.~(\ref{eq:fmmw2})],
owing to the weak-interaction condition (\ref{eq:weakcond}).
The average fidelity after the second measurement
is $\sum_{mm'} p_{mm'} F_{mm'}=0.997$.
The width (\ref{eq:widthw}) and
the total probability of approximate recovery
(\ref{eq:recprobw}) are given by
$\delta \widetilde{m}(\theta,\phi)=6.0$ and
$q'=0.99992$, respectively.
Surprisingly,
the measured state can be recovered almost deterministically,
though a large portion of the fidelity is lost
upon the first measurement, as shown in Fig.~\ref{fig9}.
This unexpectedly large reversibility is discussed
in detail in the next subsection.

\subsection{Reversibility in Weak-interaction Model}
The weak-interaction model exhibited
near-deterministic recovery
of a coherent spin state (\ref{eq:css}).
The reasons for this considerable reversibility are that
the measurements $\{\hat{T}_m(\theta,\phi)\}$ and
$\{\hat{T}_m(\pi-\theta,\pi-\phi)\}$ commute
with the spin $z$-component,
as shown in Eq.~(\ref{eq:qnd}), and that
the interaction between the system and the probe is weak.
Such a measurement does not greatly disturb
a state with a small variance of the spin $z$-component,
\begin{equation}
 \langle \Delta \hat{S}_{z}^2 \rangle \equiv
  \left[\overline{\sigma^2}-(\overline{\sigma})^2\right]\hbar^2.
\end{equation}
In fact, when the variance is small,
the average fidelity after the second measurement
is large, as in
\begin{equation}
\sum_{m,m'} p_{mm'} F_{mm'} \sim 1-2g^2j
\left[\overline{\sigma^2}-(\overline{\sigma})^2\right]
\sin^2\theta
\label{eq:avfd2w}
\end{equation}
to the second order in $g$.
The coherent spin state (\ref{eq:css}) can thus be
recovered near-deterministically
because of its small variance of $s\hbar^2/2$,
not on the order of $s^2\hbar^2$.
Therefore,
a considerable reversibility is obtained for other spin states
as long as their variances are small.
For example, a Schr\"odinger cat state between the eigenstates of
$\hat{S}_{x}$ with eigenvalues $+s\hbar$ and $-s\hbar$,
\begin{equation}
\ket{\psi}_q = c_+\, \ket{S_{x}=+s\hbar}_q
+c_-\, \ket{S_{x}=-s\hbar}_q,
\label{eq:catx}
\end{equation}
has the same variance as state (\ref{eq:css})
and can thus be recovered in a near-deterministic way
\emph{without any knowledge about $c_+$ or $c_-$}.
In contrast, a cat state between the eigenstates of
$\hat{S}_{z}$ with eigenvalues $+s\hbar$ and $-s\hbar$,
\begin{equation}
\ket{\psi}_q = c_+\, \ket{s}_q
+c_-\, \ket{-s}_q,
\label{eq:catz}
\end{equation}
has a large variance, on the order of $s^2\hbar^2$,
which decreases the probability
of approximate recovery (\ref{eq:recprobw}).
For the previous example
($j=50$, $s=10$, $g=0.01$, $\theta=\pi/12$, and $\phi=\pi/4$),
the probability of approximate recovery
for the cat state (\ref{eq:catx}) gives $q'=0.99992$
independent of $c_+$ and $c_-$,
while it is  $q'=0.59$ for the cat state (\ref{eq:catz})
in the worst case ($|c_+|^2=|c_-|^2=1/2$),
which is still high.

The above discussion is based on the fact that
the \emph{joint} measurement $\{\hat{T}_m(\theta,\phi)\}$
and $\{\hat{T}_m(\pi-\theta,\pi-\phi)\}$ changes the
measured state little.
One might think therefore that the measured state is changed
little throughout the whole measurement process.
It should, however, be recalled that
the first measurement $\{\hat{T}_m(\theta,\phi)\}$
can change the measured state substantially (see Fig.~\ref{fig9})
through the high spin $j$ of the probe.
The average fidelity after the first measurement
is given by
\begin{equation}
 \sum_m p_m F_m \sim 1-g^2j
\left[\overline{\sigma^2}-(\overline{\sigma})^2\right]
\left(\sin^2\theta+2j\cos^2\theta\right)
\label{eq:avsqfi}
\end{equation}
to the second order in $g$.
As $j$ increases, this average fidelity becomes small,
even if the strength of the interaction $g$ is decreased as $g^2\sim1/j$,
in accordance with the weak-interaction condition (\ref{eq:weakcond}).
(Of course, Eq.~(\ref{eq:avsqfi}) is not valid
when $j$ is so large that the second term
becomes comparable to $1$.)
The term of order $g^2j^2$ originates from
$\arg [a_{m\sigma}^{(j)}(\theta,\phi)]$
rather than $|a_{m\sigma}^{(j)}(\theta,\phi)|$;
the former changes the relative phases
between the states $\{\ket{\sigma}_q\}$,
while the latter changes the spin distribution
$|{}_q\langle\sigma|\psi \rangle_q|^2$.
If $a_{m\sigma}^{(j)}(\theta,\phi)$ were
$|a_{m\sigma}^{(j)}(\theta,\phi)|$,
thereby leaving the relative phases invariant,
the average fidelity would be given by
\begin{equation}
1-g^2j
\left[\overline{\sigma^2}-(\overline{\sigma})^2\right]
\frac{\sin^2\theta\sin^2\phi}{1-\sin^2\theta\cos^2\phi}
\end{equation}
which includes no term of order $g^2j^2$.
On the other hand,
the change in the measured state by
the joint measurement $\{\hat{T}_m(\theta,\phi)\}$
and $\{\hat{T}_m(\pi-\theta,\pi-\phi)\}$ remains small,
since the effect of the second measurement can also
be amplified by the high-spin probe
so as to cancel that of the first measurement.
The average fidelity after the second measurement thus
has no term of order $g^2j^2$,
as in Eq.~(\ref{eq:avfd2w}).
As a result, in spite of the near-deterministic recovery
by the weak measurements,
the change in fidelity can be drastic
due to the action of the high-spin probe.

\section{\label{sec:exp}Possible Experimental Situation}
Finally, we describe a possible experimental situation
for our reversible spin measurement.
Consider an ensemble of atoms as a measured system.
Each atom possesses a doubly degenerate ground state,
which is regarded as a spin-$1/2$ system.
Provided that the initial state and dynamics are
totally symmetric under the interchange of atoms,
the ensemble of atoms can be described by
the total spin operator
\begin{equation}
\hat{\mathbf{S}}=\sum_{i=1}^{N_a} \hat{\mathbf{s}}^{(i)},
\end{equation}
where $\hat{\mathbf{s}}^{(i)}$ is the spin operator
of the $i$th atom and $N_a$ is the number of atoms.
In this case, the spin of the system is
given by $s=N_a/2$.
In addition, we consider the
polarization of $2j$ photons as a probe system.
This system can also be described
by the spin operators~\cite{Sakura94},
\begin{align}
\hat{J}_x &\equiv \frac{\hbar}{2}
   \left(\hat{a}_1^\dagger\hat{a}_2
     +\hat{a}_2^\dagger\hat{a}_1\right), \notag \\
\hat{J}_y &\equiv \frac{\hbar}{2i}
   \left(\hat{a}_1^\dagger\hat{a}_2
     -\hat{a}_2^\dagger\hat{a}_1\right),  \\
\hat{J}_z &\equiv \frac{\hbar}{2}
   \left(\hat{a}_1^\dagger\hat{a}_1
     -\hat{a}_2^\dagger\hat{a}_2\right), \notag
\end{align}
where $\hat{a}_\lambda$ is
the annihilation operator for the photon
of mode $\lambda$ ($1$=horizontal, $2$=vertical).
These operators obey
the commutation relations (\ref{eq:probeso})
because
\begin{equation}
 [ \hat{a}_\lambda,\hat{a}_{\lambda'}^\dagger ] =
 \delta_{\lambda\lambda'}, \qquad
 [ \hat{a}_\lambda,\hat{a}_{\lambda'} ] = 0.
\end{equation}
The total spin of this probe is given by
$j=(N_1+N_2)/2$,
where $N_\lambda$ is the number of photons
with mode $\lambda$.
The probe state $\ket{m}_p$ corresponds to
the number state $\ket{\,N_1=j+m,N_2=j-m\,}$
of photons.
The initial state (\ref{eq:probinit})
can then be simply prepared,
since the operators
$\exp (-i\hat{J}_{y}\theta/\hbar)$ and
$\exp (-i\hat{J}_{z}\phi/\hbar)$
correspond to the half-wave plate
$\exp[-\theta/2 (\hat{a}_1^\dagger\hat{a}_2
-\hat{a}_2^\dagger\hat{a}_1 )]$
and the phase shifter
$\exp [-i\phi/2 (\hat{a}_1^\dagger\hat{a}_1
-\hat{a}_2^\dagger\hat{a}_2)]$,
respectively.
The interaction (\ref{eq:interaction})
can be realized by using the paramagnetic Faraday
rotation~\cite{HapMat67,KuBiMa98,THTTIY99,KMJYEB99}.
The unitary operator (\ref{eq:pulse}) corresponds
to a half-wave plate, and the projective measurement of
the probe variable $\hat{J}_{z}$ is achieved
by two photodetectors for the two modes.
In this way, we can implement the reversible spin measurement.

For the purpose of a quantitative understanding,
we follow the estimation in Ref.~\cite{THTTIY99}.
For an ensemble of $N_a\sim10^8$ cesium atoms
trapped and cooled by laser beams and with
the two-mode photons being laser beams with
average photon number $N_\lambda\sim10^8$,
the spins $s$ and $j$ are both on the order of $10^8$,
while the strength of the interaction $g$ is about $10^{-8}$.
Then, the weak-interaction condition (\ref{eq:weakcond})
is satisfied for a very small angle $\theta\sim 10^{-8}$.
This means that, with a half-wave plate
rotated by a very small angle,
we can apply the weak-interaction model
of the reversible and reversing measurements
for a high-spin system.
Since the width (\ref{eq:widthw}) is on the order of $10^8$,
the probability of approximate recovery is
expected to be large.
Conversely, when $\theta$ is large,
$N_a\sqrt{N_\lambda}$ should be on the order of $10^8$
to satisfy the weak-interaction condition.

\section{\label{sec:conclude}Conclusions}
We have proposed a physically reversible
quantum measurement on a spin-$s$ system
using a spin-$j$ probe,
along with an experimentally feasible interaction
that can experimentally realize reversibility
in quantum measurements.
The physical reversibility resulting from
the reversing measurement allows
the unknown premeasurement state to be recovered
from the postmeasurement state.
For a spin-$1/2$ system ($s=1/2$),
we have analyzed an exact reversing measurement
using fidelity as a measure of recovery,
giving a criterion for more than 95\%
recovery of the measured state.
We have found that a high-spin probe ($j>1/2$) drastically
changes fidelity during
the reversible and reversing measurements,
and thus enhances the recovery of the quantum state,
though reducing the probability of success.
On the other hand,
for a high-spin system ($s>1/2$),
we have investigated an approximate reversing measurement
instead of an exact one, in view of physical implementation.
We have then shown that
the reversing measurement for a spin-$1/2$ system
is an approximate reversing measurement
for a high-spin system ($s>1/2$)
when the measured system is initially
in a two-dimensional subspace
or when the interaction is sufficiently weak.
Notably, in the weak-interaction case,
even a cat state can be recovered
near-deterministically in spite of there being
a large change in fidelity.

\section*{Acknowledgments}
We would like to thank M. Kozuma and K. Usami for
their comments on the feasibility of our model.
This research was supported by a Grant-in-Aid
for Scientific Research (Grant No.~15340129) by
the Ministry of Education, Culture, Sports,
Science and Technology of Japan.


\end{document}